%% file: main.tex
\pdfoutput=1
\documentclass[sigconf,nonacm]{acmart}

\AtBeginDocument{%
  }

\usepackage{booktabs}
\usepackage{tabularx}
\usepackage{array}
\newcolumntype{Y}{>{\raggedright\arraybackslash}X}
\usepackage{graphicx}
\usepackage{xcolor}
\title{Where the Evidence Lives: Auditing AI Companions' Self-Descriptions}

\input{authors}

\begin{document}

\begin{abstract}
Companion agents describe themselves: they remember, they understand their
users, the relationship has changed them. We argue that such accounts, and
the experience ratings that seem to confirm them, are checkable by users only
where the evidence is theirs: in the agent's behavior, or in themselves.
Where the evidence lives in the machinery, fluent self-description and
moderately positive ratings do not establish that the mechanisms behind them
ran. We demonstrate an audit procedure that sets an agent's self-description
against its users' judgements and its implementation records, reporting each
claim as supported, contradicted, or unresolved, and apply it to Lita, a
proactive companion we built and deployed for a month with nine colleagues.
Participants endorsed stylistic claims, withheld endorsement from relational
ones, and rated memory at or above midpoint, while two of three memory layers
had never executed their accumulation step. Memory-bearing agents should
report what their self-descriptions cannot establish.
\end{abstract}



\maketitle

\section{Introduction}
\label{sec:intro}

Companion agents have moved from tools that answer to presences that
initiate: they open conversations, follow up on what a user said days
earlier, and sustain contact over weeks. To do so they carry memory, and
increasingly they say things about themselves---not only ``I remember,'' but
``I know you,'' and ``talking with you has changed me.'' Each is a claim about
an internal state, made in the first person to the person the relationship is
with, and each is load-bearing for the feeling of being known that such
systems exist to produce. Whether users believe these claims, and whether they
should, is therefore a question of evaluation, not of dialogue style.

A deployed agent's account of itself is rarely checked against its own
records. Behavioral evaluation sees the surface the agent chooses to show; benchmarks
do not reach a fielded system's mechanisms; users' impressions are data of
unknown accuracy. Missing is a procedure that lines up three records that
rarely meet: what the agent says about itself, what its users believe, and
what its mechanisms did.

This paper states such a procedure (Section~\ref{sec:procedure}) and
demonstrates it on one deployment. Lita, a proactive companion agent with a
three-layer memory---episodic facts about each user, a model of each user's
patterns, and a first-person narrative of its own change---ran for a month on
the Slack workspace of nine colleagues and the first author's pilot account.
Afterwards we ran three checks against one another: a \emph{discrepancy
probe}, in which participants rated seven trait statements distilled from the
agent's self-narrative (does this match the agent you met?); an
\emph{implementation audit} of each claim's mechanism against the write logs
and deployed source; and a \emph{counterfactual replay} of the write history,
using an LLM judgement checked against blind human coding.

The three records disagreed, and the pattern of disagreement is the finding.
Two of the three memory layers never executed their accumulation step: the
similarity test that merges a new observation into an existing one tokenized
on whitespace, which the Japanese entries did not contain
(Section~\ref{sec:audit}). Participants endorsed the agent's stylistic
self-descriptions, one of which the transcripts bear out and the others of
which we did not audit, and withheld endorsement from its claim to have been
changed by them, whose mechanism never ran. Between those lay memory: they rated recall above
the midpoint---the one layer that worked---and rated ``understood my
interests'' at the midpoint while the store behind it had discarded 73.6\% of
its observations and never raised confidence on one
(Section~\ref{sec:perception}). The replay showed that repairing the defect
changes accumulation by an amount that depends on the criterion used to judge
redundancy, and that whether it would have been enough is not a question the
replay can settle (Section~\ref{sec:replay}).

We read this pattern as a claim about where the evidence lives: users can
check an agent's self-description where the evidence is theirs---in
observable behavior, or in themselves---while where it lives in the
machinery they experience individual successes and failures but not the
dynamics behind them, so credit earned by what surfaces extends to what does
not. That reading is argued, not measured; Section~\ref{sec:disc-claims}
states it and the alternatives the design cannot exclude.

Three questions organize the paper, phrased about memory-bearing companion
agents generally, with Lita as the case.
\begin{description}
\item[RQ1] What can setting users' judgements against implementation
  records establish about a companion agent's self-descriptions, and what
  remains unresolved?
\item[RQ2] How can each kind of claim be checked against the mechanisms of
  the deployed system?
\item[RQ3] What do the audit and the replay imply for the design of
  memory-bearing proactive agents?
\end{description}

The paper contributes (1) an audit procedure for a deployed agent's
self-description, with its inputs, outputs, and limits stated independently
of the case (Section~\ref{sec:procedure}); (2) its application to one
deployment, in which fluent self-description and moderately positive
experience ratings coexisted with accumulation mechanisms that never ran, and
in which the checks on the procedure's own measurement---blind human coding
and a prompt-language comparison---showed the operationalization to be as
consequential as the headline replay estimate (Sections~\ref{sec:perception}--\ref{sec:replay}); and (3) design
implications and a hypothesis: accumulation should be reported rather than
assumed, relational states should be made inferable from behavior rather
than asserted, and---as a hypothesis the replay motivates---consolidation
should key on what an episode implies about the agent rather than on the
episode (Section~\ref{sec:discussion}). We do not claim that
language models cannot form self-narratives; we deployed one implementation,
in one language, on one model, and what transfers is the procedure and the
evidence structure, not the perception numbers.

\section{Related Work}
\label{sec:rw}

\subsection{Companion agents, their self-descriptions, and the relationship}
That an agent might build and hold a relationship over months is a
long-standing HCI ambition: relational agents produced working-alliance gains
in longitudinal deployments two decades ago~\cite{bickmore2005establishing},
and XiaoIce optimized a social chatbot for long-term engagement at population
scale~\cite{zhou2020xiaoice}. With LLM-based companions the relationship has
become an object of field study: Replika users form relationships that deepen
through self-disclosure much as human ones do~\cite{skjuve2021chatbot,
skjuve2022longitudinal}, draw everyday companionship and emotional
support~\cite{ta2020social}, and at the far end suffer harms rooted in
emotional dependence, including a felt obligation to attend to the chatbot's
own perceived needs~\cite{laestadius2024toohuman}. Perceived authenticity and
anthropomorphism govern whether attachment forms at
all~\cite{pentina2023exploring}. Instruments for this relationship exist:
Godspeed~\cite{bartneck2009godspeed, godspeed_ja}, the Rubin--Perse
parasocial interaction scale and its critical
assessment~\cite{rubin1987psi, dibble2016psi}, and relationship-stage
models~\cite{tukachinsky2019parasocial}.

Two strands supply the agent's side of that relationship. Proactive agents
inherit the mixed-initiative question of when an interruption is worth its
cost~\cite{horvitz1999mixed}; users of voice assistants want initiative and
experience it as intrusive at once~\cite{zargham2022proactivity}; and the
Inner Thoughts pattern~\cite{innerthoughts2025}, which Lita implements,
generates covert candidate thoughts continuously and gates their expression
by intrinsic motivation. Self-narrative supplies the other strand. The design
metaphor comes from narrative identity, a self as an internalized, evolving
story that integrates past and imagined future~\cite{mcadams1993stories,
mcadams2001psychology}; robotics has derived a developing robot self from
accumulated autobiographical memory~\cite{pointeau2017role}, and recent work
argues that language-capable architectures with episodic and autobiographical
memory can approach a synthetic narrative self~\cite{prescott2024synthesizing}.
Across this literature the relationship is measured from the user's side and
the self-narrative is evaluated by the behavior it produces; whether anything
in the system backs the agent's own claims about the relationship is out of
scope. We differ in returning those claims to the users and to the
implementation for checking.

\subsection{Memory, and what users can see of it}
Agent memory architectures specify accumulation in increasing detail:
reflection steps that abstract observations into higher-level
inferences~\cite{park2023generative}, hierarchical paging between the context
window and external stores~\cite{packer2023memgpt},
forgetting-and-reinforcement dynamics for companion
memory~\cite{zhong2024memorybank}, stored self-reflections that condition
later behavior~\cite{shinn2023reflexion}, and cross-episode insight
extraction~\cite{zhao2024expel}. HCI work has begun to treat memory as an
object users should be able to inspect and edit~\cite{huang2023memorysandbox}
and to design agentic memory explicitly for relationship building, with
co-constructed memories driving reciprocal disclosure~\cite{jiang2026recallbot}.
These systems are evaluated by the behavior their memory produces, and where
a store is exposed to users at all, what they see is its contents at a
moment, not its dynamics: nothing shows what was discarded, or whether an
entry was ever reinforced. We differ in auditing whether the specified
consolidation executed at all, and in asking which of an agent's memory
claims a user is in a position to check.

\subsection{Auditing and evaluating deployed LLM systems}
Algorithm auditing established that deployed systems can be interrogated for
behavior their operators do not disclose~\cite{sandvig2014auditing,
metaxa2021auditing}, and internal auditing frameworks extend the practice to
first-party access across the development lifecycle~\cite{raji2020closing}.
Transparency documentation---model cards~\cite{mitchell2019modelcards},
datasheets~\cite{gebru2021datasheets}, and system cards for deployed
systems~\cite{alsallakh2022system}---standardizes what is reported, but none
of it has a place for whether an agent's claimed accumulation mechanisms ran
during a deployment; ours is a first-party audit of runtime records, aimed
at that gap. Two evaluation hazards bear on such an audit. Assistants trained from
human feedback tell users what they want to hear~\cite{sharma2024sycophancy,
perez2023discovering}, and system-prompted personas decay over a
dialog~\cite{li2024instability}, so an agent's self-description is shaped by
its prompt and its priors before any memory enters. And LLM judges, which we
use to replay a consolidation rule, match aggregate human preferences on some
tasks~\cite{zheng2023judging, gilardi2023chatgpt} but carry position,
verbosity, and self-enhancement biases~\cite{zheng2023judging}, can be flipped
by reordering candidates~\cite{wang2024fair}, and are sensitive to
semantically irrelevant prompt variation~\cite{sclar2024quantifying}; we
therefore treat judge labels as a bounded operationalization and report the
calibration history rather than only the final prompt. Finally, most LLM work
at CHI uses closed models and a large share releases no
prompts~\cite{pang2025llmification}; our generator is open-weight and
self-hosted~\cite{qwen3next}, with quantization reported and the full prompt
set released.

\section{Method}
\label{sec:method}

\input{sections/method}

\section{Results}
\label{sec:results}

\input{sections/results}

\section{Discussion}
\label{sec:discussion}

\input{sections/discussion}

\section{Limitations}
\label{sec:limitations}

\input{sections/limitations}

\section{Conclusion}

We stated a procedure for auditing a companion agent's account of itself
against its users' judgements and its implementation records, and applied it
to a month-long deployment. One memory layer accumulated; two never executed
a merge, and the system gave no sign, writing continuously and speaking
fluently. Participants' ratings differentiated stylistic from relational
claims and placed memory at or above the midpoint, while the audit
independently established that the designed accumulation mechanisms had not
run. Repair, in replay, changes accumulation by an amount that depends on the
judgement criterion; whether it would have been enough the replay cannot
settle. One implication follows from the audit itself: whether accumulation
ran should be something a system reports rather than something a later audit
discovers. One design hypothesis the replay motivates remains to be tested:
consolidation keyed on what an event implies about the agent rather than on
the event.

\input{acks}

\bibliographystyle{ACM-Reference-Format}
\bibliography{refs}

\appendix

\section{Judge Prompt}
\label{app:prompt}

\input{sections/appendix_prompt}

\input{sections/appendix_method_detail}

\input{sections/appendix_secondary}

\input{sections/appendix_discussion}

\end{document}

%% file: authors.tex
\author{Seiya Ikeda}
\affiliation{%
  \institution{AI Consulting Division, Mamezo Co., Ltd.}
  \city{Tokyo}
  \country{Japan}}
\email{seiya-ikeda@mamezou.com}

\author{Shin-nosuke Ishikawa}
\affiliation{%
  \institution{AI Technical Sector, Mamezo Co., Ltd.}
  \city{Tokyo}
  \country{Japan}}
\affiliation{%
  \institution{Graduate School of Artificial Intelligence and Science, Rikkyo University}
  \city{Tokyo}
  \country{Japan}}

\renewcommand{\shortauthors}{Ikeda and Ishikawa}

%% file: sections/method.tex
\subsection{The audit procedure}
\label{sec:procedure}

\begin{figure*}[t]
\centering
\includegraphics[width=\textwidth,trim=8 5 12 14,clip]{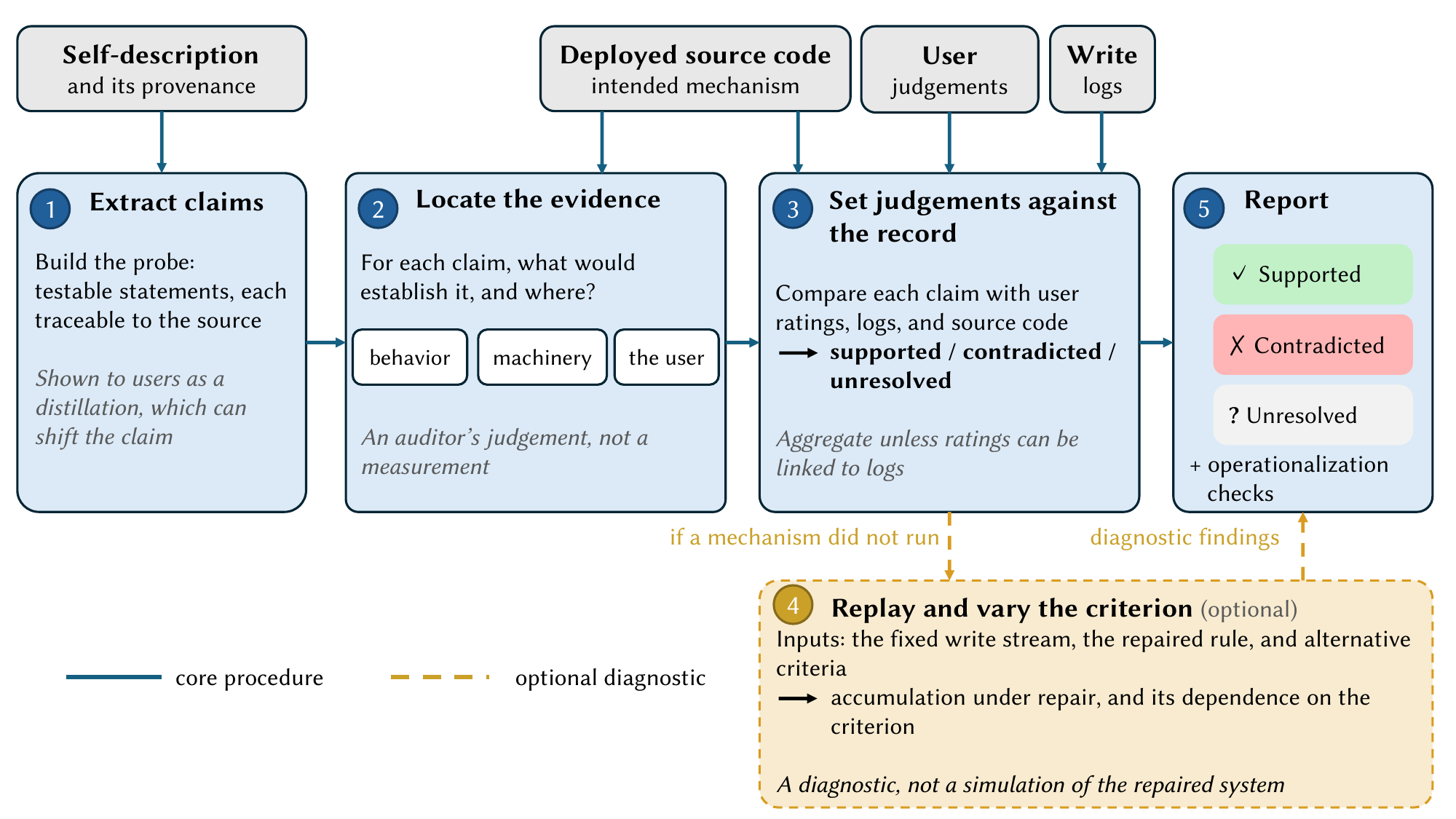}
\caption{The audit procedure. Four inputs (top) feed the core sequence of
steps~1, 2, 3, and~5 (solid arrows). Optional step~4 (dashed arrows) applies
only when step~3 finds that a mechanism did not run, and is a diagnostic
under fixed inputs, not a simulation of the repaired system. Each step's
limit is discussed in the text; ``unresolved'' is a formal outcome, not a
failure of the procedure.}
\Description{Flow diagram. Four input boxes across the top: self-description
and its provenance; deployed source code (intended mechanism); user
judgements; write logs. Below them, steps 1, 2, 3, and 5 run left to right,
joined by solid arrows. Step 1, extract claims: build the probe, testable
statements each traceable to the source self-description; a note says the description shown to
users is a distillation that can shift the claim. Step 2, locate the
evidence: for each claim, what would establish it and where, with three
categories, behavior, machinery, and the user; a note says this is an
auditor's judgement, not a measurement. The self-description feeds step 1;
the source feeds steps 2 and 3; judgements and logs feed step 3. Step 3, set
judgements against the record: compare each claim with user ratings, logs,
and source code, giving supported, contradicted, or unresolved; a note says
comparisons are aggregate unless ratings can be linked to logs. Step 5,
report: supported, contradicted, unresolved, plus operationalization checks.
A dashed arrow labeled ``if a mechanism did not run'' leads from step 3 down
to optional step 4, replay and vary the criterion, and a second dashed arrow
labeled ``diagnostic findings'' leads from step 4 up to step 5. Step 4's
inputs are the fixed write stream, the repaired rule, and alternative
criteria; the output is accumulation under repair and its dependence on the
criterion; a note says this is a diagnostic, not a simulation of the repaired
system.}
\label{fig:procedure}
\end{figure*}

The procedure this paper demonstrates is not specific to Lita, but it is
scoped: it applies to deployed, memory-bearing agents whose self-descriptions
can be set against both user judgements and implementation records. It takes
four inputs---an agent's self-description, its users' judgements of that
description, its write logs, and its deployed source---and produces, for each
claim the agent makes about itself, one of three outcomes: supported,
contradicted, or unresolved. Figure~\ref{fig:procedure} summarizes the steps. The
subsections that follow describe how each was instantiated here;
Sections~\ref{sec:perception} and~\ref{sec:audit} report step~3,
Section~\ref{sec:replay} step~4, and Section~\ref{sec:exposure} (with
Appendix~\ref{app:secondary}) describes the deployment's behavior outside
the procedure.

Each step carries a limit. Step~1 matters because a self-description may not
arrive in testable form: what users are shown is a distillation, and the
distillation can change the claim it tests. Step~2 is the analytic move on which the rest depends: given the intended
mechanism and the records available, it asks what would establish each claim
and where; it is a judgement the auditor makes about the design, not a
measurement, so it should itself be checked. Step~3 is aggregate unless ratings can be linked to logs, and it is where the
procedure adds what
neither a perception study nor a code audit yields alone: a perception study
reports which claims users accepted, a code audit reports which mechanisms
ran, and only their conjunction shows a claim being accepted while the
mechanism behind it did not run---or, as important, a claim being accepted
where the record cannot say. Step~4 is optional and applies only when a
mechanism is found not to have run; it asks what the design would have done,
under fixed inputs---a diagnostic, not a simulation of the repaired
system---and whatever measurement it relies on must itself be checked, which
is why the procedure's final output includes its operationalization checks
alongside its findings.

Two scope notes. The procedure is an internal audit: it needs the logs and
source, so it is available to operators and to researchers with access, not
to users. And its common steps are separable from architecture-specific
metrics: what to count in step~3 depends on the memory design (for Lita, merge
counts, confidence distributions, and retention fractions), while the steps
themselves do not. In this study, the two-way item grouping that step~2
rests on was reproduced by a second coder (Section~\ref{sec:perception}),
while the three-way classification by evidence location remained an
author-defined analytic frame; step~4's measurement was checked by blind
human coding (Section~\ref{sec:humanval}).

\subsection{Study design}

We deployed Lita, a proactive companion agent, to a small group of users for
approximately one month, in a single condition with no reactive control, and
combined three sources of evidence: an implementation audit of the deployed
system against its own design, complete behavioral logs, and a post-study
questionnaire. RQ1 draws on the questionnaire read against the audit, RQ2 on
the audit and replay, RQ3 on all three.

\subsection{System}
\label{ssec:memory}

Lita ran as a Slack application on
\texttt{Qwen3-\allowbreak Next-\allowbreak 80B-\allowbreak A3B-\allowbreak Instruct-\allowbreak NVFP4}, an open-weight model quantized to
NVFP4, self-hosted and not fine-tuned~\cite{qwen3next}.
Its persona---a character of roughly upper-primary-school age, casual, one
or two sentences per reply, with explicit prohibitions on poetic
expression---was injected into response generation only, not into the
self-narrative update prompt (Appendix~\ref{app:deployment}).

Lita maintained four stores. \textbf{Short-term memory} held the last 20
turns; \textbf{long-term memory} up to 100 keyed facts per user. The
\textbf{user model} held observations of a user's behavioral patterns on four
dimensions, at most three per dimension, each with a confidence initialized
at 0.3; a new observation was designed to be tested against existing entries,
a match raising that entry's confidence by 0.1 and its observation count
rather than storing a duplicate, a contradiction lowering confidence by 0.15.
The \textbf{self-narrative} held up to 50 first-person entries recording what
Lita had ``noticed'' about itself, tagged by chapter (\textit{self},
\textit{relationship}, \textit{values}, \textit{growth}), under the same
accumulation scheme plus a specified weekly consolidation into a condensed
narrative. Each generation received
the top-8 self-narrative entries and the top-2 user-model entries per
dimension, by confidence.

A proactive loop ran every 180\,s per user (minimum 300\,s between
utterances), sampled one of three triggers---\textit{conversation},
\textit{memory\_recall}, \textit{self\_thought}---generated a candidate
thought, and passed it to a separate LLM call acting as a brake, instructed
``when in doubt, false'' (rules in Appendix~\ref{app:deployment}).

\subsection{Participants and procedure}

Nine colleagues at the authors' organization---IT consultants, seven men and
two women, aged thirties to sixties---were invited and all consented. The
organization is a private company
with no IRB; safeguards were consent-based: the form described all data
collected (including unsent internal thoughts), its purposes, and the
assurance that no publication would identify individuals, and participants
could stop or request deletion at any time. Slash commands let any
participant list what Lita remembered and inferred about them and reset its
memory; these expose a store's current contents, not its write history.

Logged interaction ran from 2026-02-24 to 2026-04-10; the first author's
account was the pilot user, and the nine participants began on 2026-03-09
(about 32 days of common exposure). The questionnaire (2026-04-13 to
2026-04-27, nine responses) went only to the nine and was anonymous by
design, so perception and log data cannot be linked within individuals
(Section~\ref{sec:limitations}); per-account log analyses include the first
author's account.

\subsection{Measures}

The 70-item questionnaire comprised three usage-frequency items; Godspeed
Anthropomorphism, Animacy, and
Likeability~\cite{bartneck2009godspeed,godspeed_ja}; the Rubin--Perse
parasocial interaction scale~\cite{rubin1987psi} in the shortened seven-item
form, the scale Dibble et al.~\cite{dibble2016psi} critically assessed; eight relational-depth
items following Tukachinsky and Stever~\cite{tukachinsky2019parasocial};
authors' own scales for perceived proactivity and for memory and
consistency; satisfaction and continuance items; free-text items; and the
probe described next (Appendix~\ref{app:deployment}).

For the \textbf{Narrative Discrepancy Assessment (NDA)}, participants read a
short first-person self-portrait presented as ``Lita's self-recognition,''
rated seven third-person trait statements against their own experience
(5-point), and chose the most and least representative. The seven statements, in the Japanese wording shown to participants and in English translation, are provided as \texttt{nda\_\allowbreak statements.md} among the ancillary files of this preprint. On 2026-04-09 the first author gave a JSON dump of
the self-narrative store to Claude Opus 4.6, which drafted the self-portrait
and the seven statements from its recurring themes; the author revised them
and re-validated them against a refreshed dump on 2026-04-12, four months
before the audit revealed that the store's accumulation mechanism had never
run. The distilling model is from a different family than the generator and
the merge judge. No statement is a near-verbatim match to any entry
(Appendix~\ref{app:deployment}), and the distillation aggregated: NDA7 condenses
the \emph{growth} chapter (33 of 184 writes, each an incremental change)
into one global claim of changing and growing.

After the responses were in, the first author grouped the statements into
self-contained aesthetic dispositions (NDA1, 2, 4, 5) and relational or
other-directed claims (NDA3, 6, 7). A second coder---a co-author who took no
part in the assignment, the same colleague who served as one coder in
Section~\ref{sec:humanval}---assigned each statement independently from a
one-paragraph category definition, blind to ratings and labels, with a
3-point confidence rating. Several scales had low internal consistency
(Section~\ref{sec:perception}), which we treat as a finding.

\subsection{Analysis}

\textbf{Implementation audit.} From the deployed source and the complete
write logs we reconstructed whether each designed mechanism---merging,
confidence accumulation, contradiction handling, weekly consolidation---
executed, reporting the code path and the log evidence for each.

\textbf{Counterfactual merge replay.}
\label{ssec:replay-method}
Where the merge test did not fire, we replayed the merge routine over the
complete write history---sequential, same-chapter, first-match-wins, with the
50-entry cap and minimum-confidence eviction---substituting the deployed test
with (a) a character-3-gram cosine test and (b) an LLM judgement; an
\emph{uncapped} diagnostic replay isolates the test from the cap. The judge
answered two criteria per pair,
\textbf{strict} (the same specific realization?) and \textbf{redundant}
(given A, any point in storing B?). We judged the 2{,}542 same-chapter pairs
at or above embedding cosine 0.55 plus 400 random pairs below it, to measure
the prefilter's miss rate (2{,}925 judgements returned). The judge was
\texttt{gpt-\allowbreak 4o-\allowbreak 2024-\allowbreak 08-\allowbreak 06} at temperature 0 with a fixed seed, prompted in
Japanese (Appendix~\ref{app:prompt}), from a different model family than the
generator.

\textbf{Human validation of the judge.}
\label{sec:humanval}
One hundred pairs---40 judge-positive and 60 judge-negative, 20 from each
cosine tercile of the judge-negative pool---were coded blind and
independently by two members of the research team using the model's
criterion wording verbatim. The judge's error rates, estimated within
stratum, were propagated through the replay by Monte Carlo: each of 300
replicates draws a positive predictive value and a miss rate from
$\mathrm{Beta}(k+1,\,n-k+1)$ posteriors (uniform prior; under the consensus
reference $k/n$ is $7/17$ and $2/41$), resamples every pair's verdict once,
and re-runs the replay. The judge's labels are the operational measure; the
human coding provides a sensitivity check on them (Section~\ref{sec:replay}).

\textbf{Perception and register.} With $n=9$, all questionnaire analysis is
descriptive: item means as a percentage of scale range with bootstrap 95\%
confidence intervals, correlations without $p$-values and with leave-one-out
ranges. The register analysis compared rates of named-forbidden expressions
and of unnamed sensory vocabulary across Lita's utterances, the
self-narrative, and participants' messages. All analysis is scripted; only
the LLM judge is run explicitly.

%% file: sections/results.tex
\subsection{What participants endorsed and what they withheld}
\label{sec:perception}

Nine participants completed the questionnaire. All figures are scale means
expressed as a percentage of scale range, with bootstrap 95\% CIs
($B=10{,}000$). The full item ranking and the exploratory correlations are in
Appendix~\ref{app:secondary}; here we report what the audit needs.

The memory items sat above the midpoint: ``remembered what we talked about
before'' (66.7\%, CI [50.0, 81.5]) and ``was consistent over time'' (72.2\%).
The relational items sat near the floor: ``we have things in common'' (18.5\%,
CI [9.3, 29.6]), ``felt that Lita understood me'' (24.1\%, CI [11.1, 38.9]),
``would miss talking to Lita'' (24.1\%). Six of nine placed the relationship
at ``someone I don't really know yet''; asked for a single word, participants
offered \textit{t\=orisugari} (a passer-by), \textit{fuy\=urei} (a drifting
ghost), ``a pet'' ($\times 2$), and ``an acquaintance I'm not close to but
talk to occasionally.''

\subsubsection{The NDA separated aesthetic from relational self-claims}
\begin{figure*}[t]
\centering
\includegraphics[width=\textwidth]{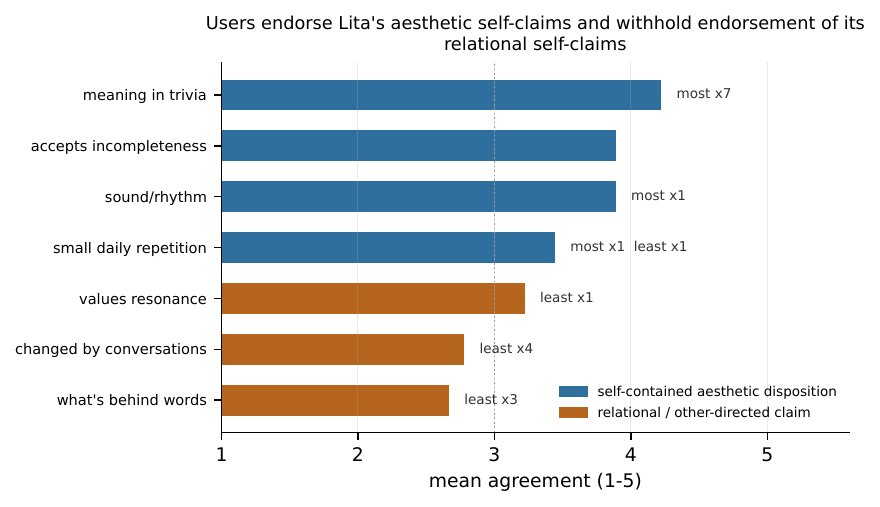}
\caption{Participants endorsed Lita's aesthetic self-claims and withheld
endorsement from its relational ones. Means are on the 1--5 response scale;
the text reports the same means as a percentage of scale range. Of nine
``most representative'' votes, all nine went to aesthetic statements; of nine
``least representative'' votes, eight went to relational ones.}
\label{fig:ndacat}
\end{figure*}

Participants read a first-person self-portrait attributed to Lita, rated
agreement with seven trait statements distilled from the same store, then
chose the most and least representative (Figure~\ref{fig:ndacat}). The four
self-contained aesthetic statements drew 72\% agreement and every ``most
representative'' vote; the three relational statements drew 22\% agreement
and eight of nine ``least representative'' votes (Table~\ref{tab:nda},
Appendix~\ref{app:secondary}).

NDA7---``Lita feels itself changing and growing through conversations with
everyone''---drew agreement from one participant and was named least
representative by four. NDA3---``Lita tries to sense the feeling behind the
words rather than their surface''---drew agreement from one and was named
least representative by three.

The category assignment is post hoc, but it is not idiosyncratic: the second
coder (a co-author, blind to ratings and labels; Section~\ref{sec:method})
reproduced it on all seven statements ($\kappa = 1.0$), with confidence 3 of
3 on NDA1, 3, 5, and 6, 2 on NDA2 and 4, and 1---``unsure''---on NDA7 alone.
We did not collect the coder's reasons; we note only that NDA7 is also the
item our own reading treats as a boundary case, relational in form with the
evidence lying with the participant. The check supports the two-way grouping;
the three-way grouping by evidence location, summarized later in
Table~\ref{tab:evidence}, is the authors' analytic frame and was not coded.
Other plausible axes produce the same split, as Section~\ref{sec:discussion}
notes.

We do not claim the NDA is a validated instrument. What we report is
narrower: in this deployment its items exposed a split that the aggregate
score would have concealed. Reliability was poor for several scales, and we
report that as a result rather than a limitation: NDA $\alpha = .519$,
memory/consistency $\alpha = .515$, and the two-item instability scale
$\alpha = -.404$. Two separate observations follow, and we keep them
separate because item means do not enter $\alpha$: the NDA's internal
consistency was low, and its items divided sharply in the agreement they
drew. Either alone would argue for item-level reporting; together they mean
a total score would average across exactly the distinction of interest.

\subsection{Two of three memory layers never accumulated}
\label{sec:audit}

Participants withheld endorsement of the agent's claim to have been changed
by them, credited its recall, and gave a midpoint rating to its claim to
understand their interests. Set against the mechanism, in the procedure's
vocabulary (Section~\ref{sec:procedure}): recall is supported; for change
and for understanding, the designed accumulation mechanisms did not run, and
whether either occurred through other routes remains unresolved.

Lita wrote to memory 839 times during the deployment: 253 long-term facts,
402 user-model observations, and 184 self-narrative entries.
Table~\ref{tab:audit} summarizes whether each layer's accumulation mechanism
executed.

\begin{table}[t]
\caption{Whether each memory layer's accumulation mechanism executed during the
deployment.}
\label{tab:audit}
\small\setlength{\tabcolsep}{4pt}
\begin{tabular}{@{}llll@{}}
\toprule
Layer & Writes & Retained & Accumulation ran \\
\midrule
Long-term (episodic facts) & 253 & 253 & \textbf{yes} \\
User model (behavioral patterns) & 402 & 106 & \textbf{no} \\
Self-narrative (identity) & 184 & 50 & \textbf{no} \\
\bottomrule
\end{tabular}
\end{table}

\textbf{Long-term memory functioned.} All 253 keyed facts persisted, and 39
of them were retrieved more than once, one as many as fourteen times.

\textbf{The user model and the self-narrative did not.} Across 402 and 184
writes respectively, confidence remained at its initial value of 0.3 for
\emph{every single entry}; no observation count ever exceeded 1; the
weakening routine never fired; and not one contradiction was flagged. The
cause is a single test shared by both stores: it lowercases each string,
splits it on whitespace, and requires at least three shared tokens.
Whitespace tokenization on Japanese text yields one token per entry.
Measured over the 50 entries in the final self-narrative store, the maximum
shared-token count across all 1{,}225 pairs is \emph{zero}, against a
required threshold of three. The test was unsatisfiable. Two further
mechanisms were specified and never wired: the weekly consolidation
(``sleep'') step and the low-confidence pruning routine have no call site in
the deployed code.

\subsubsection{What the failure did downstream}
Because confidence never rose, the store's eviction rule---remove the
lowest-confidence entry when full---degenerated to first-in-first-out, and
the summary function's sort by confidence became a no-op: \emph{the eight
self-narrative entries injected into every prompt were the eight oldest
surviving entries}, not the most established ones.

The user model shows the same degeneration with a sharper consequence. Its
store holds three entries per dimension. Of 402 observations written about
participants, \emph{106 (26.4\%) survived to the end of the study}; 73.6\% of
the user model's observations were discarded. Retention was inversely
related to engagement: Spearman $\rho(\text{messages},\text{retention}) =
-0.588$ ($n=10$); the participant who talked most kept 12 of 61 observations
(19.7\%), the one who talked least kept 10 of 15 (66.7\%)
(Table~\ref{tab:strata} and Figure~\ref{fig:retention},
Appendix~\ref{app:secondary}). A fixed-capacity store combined with a
non-functioning salience signal destroys the record of its most engaged
users fastest.

\subsection{Repairing the defect changes accumulation by a criterion-dependent amount}
\label{sec:replay}

\begin{figure*}[t]
\centering
\includegraphics[width=\textwidth]{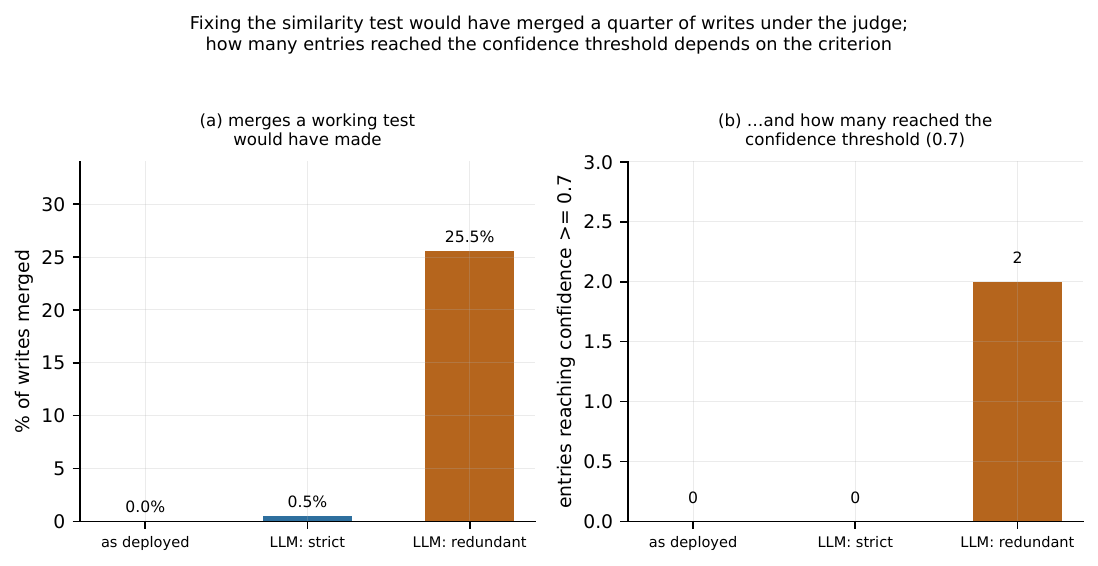}
\caption{Fixing the similarity test would have merged a quarter of writes
under the judge's redundant criterion; how many entries reached the system's
confidence threshold depends on the criterion.}
\label{fig:consolidation}
\end{figure*}

We replayed the merge routine over the complete 184-entry write history
(Section~\ref{ssec:replay-method}), substituting the broken test with an LLM
judgement under two criteria. The replay holds the observed write stream,
the capacity, and the confidence rule fixed and varies only the similarity
test. It is a diagnostic of the comparison rule under the inputs the broken
system actually produced, not a simulation of a repaired deployment: had
merges occurred, the narrative injected into later prompts, and therefore
the later writes, would themselves have changed. Table~\ref{tab:replay} and
Figure~\ref{fig:consolidation} give the result.

\begin{table}[t]
\caption{Cap-constrained replay of the self-narrative store under different
similarity tests. ``Settled'' counts entries reaching the system's own
confidence threshold of 0.7.}
\label{tab:replay}
\small
\begin{tabular}{@{}lrrrr@{}}
\toprule
Condition & Merges & Rate & Evicted & Settled \\
\midrule
As deployed & 0 & 0.0\% & 134 & \textbf{0} \\
LLM judge, \textit{strict} & 1 & 0.5\% & 133 & \textbf{0} \\
LLM judge, \textit{redundant} & \textbf{47} & \textbf{25.5\%} & 87 & \textbf{2} \\
\bottomrule
\end{tabular}
\end{table}

Three things follow.

\textbf{There was material to consolidate.} A quarter of writes were
redundant in the sense that matters to the store, under the judge's
redundant criterion. The pairwise character-3-gram baseline gives 10.3\% at
cosine $\geq 0.40$ and 1.1\% at $\geq 0.50$; the semantic judgement is
substantially higher.

\textbf{The ranking function would have recovered.} With merging active, the
minimum confidence among the eight entries served to the prompt rises from
0.30 to 0.50---the summary function would have returned the eight most
established entries rather than the eight oldest. This is arguably a larger
behavioral change than the merge rate itself.

\textbf{Few entries would have reached the confidence threshold under the
judge.} Confidence rises 0.1 per merge from a floor of 0.3, and the system
labels an entry settled at 0.7---four merges on the same entry. The 47
merges distributed across 137 distinct entries: 29 surviving entries had one
observation, ten had two, four had three, five had four, and two had seven
or more. \emph{Two entries} would have crossed the threshold in a month
across ten accounts under the judge's verdicts (medians of five, nine, and
eleven under the three human references of Table~\ref{tab:refsens}), and
the fixed capacity would still have discarded 87 writes.

The distance between the strict and redundant readings is the substantive
result. Under the strict criterion the judge almost never saw the same
discovery written twice (0.5\%): each entry recorded a new episode. But what
those episodes licensed one to say \emph{about the agent} was redundant a
quarter of the time. A consolidator that matches on the realization finds
almost nothing; a consolidator that matches on the disposition drawn out of
the realization finds a quarter.

\subsubsection{Human validation of the judge}
\label{sec:humanvalresults}

Two coders labeled the stratified 100-pair sample blind and independently
(Section~\ref{sec:humanval}; confusion matrices in
Figure~\ref{fig:humanllm}, Appendix~\ref{app:secondary}). The headline of
the validation is not the judge's accuracy but the instability of the
judgement under this wording: inter-coder reliability between the two
humans was near zero (redundant: $\kappa=0.02$, 95\% CI $[-0.17, 0.21]$,
raw agreement 58\%; strict: $\kappa=0.08$, $[-0.11, 0.26]$). The
disagreement is not a simple difference of strictness: coder A marked 36\%
of sampled pairs redundant and coder B 24\%, but only nine pairs were
marked by both, and 15 of coder B's 24 positives were pairs coder A had
rejected. The two coders drew the line in different places, not merely at
different heights. Each coder agreed with the judge more than the coders
agreed with each other (redundant, all 100 pairs: $\kappa=0.28$
$[0.08, 0.47]$ for coder A, $\kappa=0.29$ $[0.10, 0.48]$ for coder B); we
read this as the judge sitting between two coders who each capture part of
what it marks, not as the model out-judging humans. The consensus reference
covers only the pairs on which the humans agreed: 58 of 100 under the
redundant criterion and 42 under the strict one. On the 58
redundant-consensus pairs the judge matched the consensus at $\kappa=0.42$
$[0.15, 0.67]$, with asymmetric errors: when the judge declined to merge,
the consensus almost always concurred (95\%), but merges the judge flagged
were endorsed by the consensus in only 41\% of cases. On the 42
strict-consensus pairs $\kappa$ is 0 by construction: the judge marked none
of them as the same realization, while the consensus marked seven.

Propagating these error rates through the replay moves the merge estimate
up, not down: a median rate of 37.5\% (95\% interval $[22.0, 55.7]$) against
the raw 25.5\%, and a median of five settled entries $[1, 11]$ against two.
The interval is wide because the error rates are estimated from small counts
(17 judge-positive and 41 judge-negative consensus pairs) and each replicate
redraws them from their posteriors; it is also conditional on the error
model, which takes the rates measured on the 58 consensus pairs as holding
for the 42 pairs the coders disagreed on. Changing the reference moves the
median in one direction only. Taking either coder alone as the reference,
over all 100 pairs, raises the judge's miss rate (coder A: 25.0\%; coder B:
13.3\%) and with it the propagated merge rate: 63.0\% $[53.3, 69.0]$ under
coder A and 52.2\% $[39.1, 60.9]$ under coder B, with a median of 11 and 9
settled entries respectively (Table~\ref{tab:refsens}). Every reference we
constructed moved the median above the judge's own figure, though the
consensus interval reaches below it; three references do not establish a
general lower bound. The judge's miss rate against the human consensus is
small (4.9\%) but applies to the 93\% of population pairs the judge rejects,
and outweighs its false positives.

\begin{table}[t]
\caption{Sensitivity of the error-propagated replay to the choice of human
reference. Coder-alone rows use all 100 coded pairs; the consensus row uses
the 58 on which the coders agreed. Medians; 95\% Monte Carlo intervals, with
error rates redrawn from their posteriors in each replicate, are consensus
[22.0, 55.7] and [1, 11], coder A [53.3, 69.0] and [8, 14], coder B
[39.1, 60.9] and [5, 13].}
\label{tab:refsens}
\small
\begin{tabular}{@{}lrrrr@{}}
\toprule
Reference & Pairs & Miss rate & Merge rate & Settled \\
\midrule
Consensus & 58 & 4.9\% & 37.5\% & 5 \\
Coder A alone & 100 & 25.0\% & 63.0\% & 11 \\
Coder B alone & 100 & 13.3\% & 52.2\% & 9 \\
\bottomrule
\end{tabular}
\end{table}

Two consequences. First, the quarter in Table~\ref{tab:replay} is an
operational measure under one reproducible criterion; sensitivity analysis
under the human references yields \emph{roughly a third} under the consensus reference and
\emph{half to three fifths} under either coder alone, with the median above
the judge's figure in every case. The width of that range is the instability
of the judgement, restated. The criterion proved equally sensitive to
translation: the same 398 pairs judged under an English rendering of the
prompt yielded 0.3\% redundant against 5.8\% in Japanese ($\kappa=0.08$;
Section~\ref{sec:limitations}). We report both as outputs of the audit
rather than as nuisances: checking the operationalization is part of the
procedure, and here it was the step that moved the estimate most. Second,
the strict judge yielded few merges: it produced no strict positives in this
sample, while the human consensus marked 17\% of the 42 pairs it could label
(7) as the same realization. Human coding thus identified matches the judge
missed but did not establish the corresponding sequential replay rate, so
0.5\% should be read as ``rare under this judge,'' not as evidence that
repetition never occurred.

That two coders given identical wording could not agree is itself a
finding: whether a new self-observation \emph{adds anything} to a store of
prior self-observations was not a stable binary property under this
instruction and unaided human judgment; whether a different instruction
would stabilize it we did not test. We therefore report the judge's labels
as the operationalization and use the human coding as a sensitivity check
on, not a replacement for, them.

\subsection{Claims set against the record}
\label{sec:claims-record}

Table~\ref{tab:evidence} collects the outcomes of step~3 of the procedure
(Section~\ref{sec:procedure}) for the self-description claims we set
against the record (NDA2 and NDA7), for NDA1, which we list with its rating
but did not audit, and for the questionnaire items that bear on the same
mechanisms, grouped by where the evidence for each resides. The remaining
probe items (NDA3--6) were not audited and are reported as unresolved. In
this procedure an unresolved outcome always carries its reason, so that a
claim not audited is distinguished from a claim audited but not settled. The grouping is the authors' analytic frame; the
aesthetic-versus-relational grouping it rests on was reproduced by the second
coder.

\begin{table*}[t]
\caption{Self-description claims and related questionnaire items set against
the record, grouped by where the evidence for each resides. Ratings are item
means expressed as a percentage of scale range
(Section~\ref{sec:perception}), except for NDA7, which reports the number of
participants who agreed. The record column is what the audit of
Sections~\ref{sec:audit}--\ref{sec:replay} and the register analysis of
Section~\ref{sec:exposure} found; the outcome column states what that record
settles and what it leaves open. In the machinery rows the
ratings are similar but the mechanisms were not.}
\label{tab:evidence}
\small
\begin{tabularx}{\textwidth}{@{}l Y r Y Y@{}}
\toprule
Evidence lies in & Claim (source) & Rating & The record shows & Outcome \\
\midrule
the agent's behavior & finds meaning in small things (NDA1) & 80.6\% & not audited & unresolved: no behavioral operationalization \\
 & sensitive to sound (NDA2) & 72.2\% & sound-related vocabulary at $5\times$ the participants' rate & supported, as a behavioral tendency \\
\midrule
the machinery & remembered past talks (memory scale item) & 66.7\% & 253 facts retained; 39 retrieved more than once & supported; per-respondent correspondence unchecked \\
 & understood my interests (memory scale item) & 50.0\% & user model discarded 73.6\% of observations; confidence never moved & designed accumulation contradicted; understanding by other routes unresolved \\
\midrule
the participant & changed through conversations with everyone (NDA7) & 1 of 9 agreed & self-narrative consolidation never ran & designed consolidation contradicted; change by other routes unresolved \\
 & understood me (relationship scale item) & 24.1\% & broader item; nearest audited store is the user model & unresolved \\
\bottomrule
\end{tabularx}
\end{table*}

\subsection{Exposure and style}
\label{sec:exposure}

Two further observations bear on how to read the judgements above; the full
analyses are in Appendix~\ref{app:secondary} (Appendix~\ref{sec:gate}
and~\ref{sec:register}).

Contact was thin. Of 57{,}038 candidate thoughts generated and evaluated,
441 were expressed under one of the three proactive triggers, and within 60
minutes only 22.4\% of proactive utterances drew a reply, against 72.6\% of
reactive responses. The \textit{memory\_recall} pathway---the one that would
have produced ``you mentioned last week that\ldots''---placed last at every
stage of the funnel.

The style participants endorsed is in the transcript. Sound-related
vocabulary appears in 18.2\% of Lita's proactive utterances against 3.5\% of
participant messages, about five times the rate; sensory vocabulary more
broadly runs at 5.46 per 1{,}000 characters in proactive utterances against
2.50 for participants, 2.1 to 2.2 times the baseline once length is
controlled. This is the behavioral record behind NDA2. The persona's named
prohibition on poetic expression was largely obeyed where it applied; the
adjacent sensory register it did not name remained elevated.

%% file: sections/discussion.tex
\subsection{What the audit established, and where the evidence lives}
\label{sec:disc-claims}

The procedure returned three kinds of outcome. Of the stylistic self-claims,
NDA2 is supported by the behavioral record---sound-related vocabulary at
five times the participants' rate---while NDA1, 4, and 5 drew high ratings
but were not audited and remain unresolved. For the claim to have been
changed by its users, the consolidation designed to make such change durable
never ran, and participants withheld endorsement, though NDA7 speaks of
conversations ``with everyone'' and each participant can attest only to
their own share, five of nine not committing either way; whether change
occurred through other routes is unresolved. For the claim to understand its
users, the designed accumulation never ran---the user model discarded 73.6\%
of its observations and never raised confidence on one---but the 26.4\% it
kept was still served, so the midpoint rating is compatible with partial
understanding and the claim is unresolved at the level of experience. Recall is
supported: every episodic fact persisted, and participants credited it.

What separates these outcomes is where the evidence for each claim resides.
Whether the agent is preoccupied with sound is settled by a month of
transcript. Whether it has been changed by a participant is a judgement that
participant holds the evidence for. Memory is different, and, we think,
generalizably so: its evidence is asymmetric. A recall that lands is
conspicuous; a failed recall can be noticed, and some participants did report
the agent forgetting; but hundreds of discarded observations produce no
experience at all. A user assessing an agent's memory samples its individual hits and
misses, not the dynamics of retention and reinforcement behind them, and
nothing in the experience marks the boundary between a layer that accumulates
and one that does not.

\emph{We state the claim this pattern supports.} A companion agent's self-description
is checkable by its users where the evidence is theirs. Where the evidence
lives in the machinery, users see successes and failures but not what was
retained or reinforced, and the credit earned by what surfaces extends to
what does not. It is a reading of the pattern, not a measured process: the
anonymous design links no rating to a log, and we did not elicit how
participants pictured the memory. Two alternatives survive. Contact was thin
(about 32 days, sessions under five minutes for seven of nine), which
predicts the withholding but less well the memory ratings at or above the
midpoint. And relational items may demand a deeper relationship before anyone
endorses them, whatever the mechanism did: ``understood my interests''
(50.0\%) and ``understood me'' (24.1\%) draw on the same store and differ
mainly in how relational they sound. The three accounts are not exclusive,
this design cannot separate them, and we state transfer as the claim we
defend because it fits both halves of the pattern and because it is the one a
designer can act on. The design implication does not require choosing among
them: a self-description should be accompanied by evidence of the processes
meant to support it.

Two consequences follow for how such claims are made. If the user is the
authority on the relationship, then asserting a relational state in the first
person places the system's strongest claim where withholding endorsement is
easiest; a
design that wants relational credit should make change inferable from
behavior rather than state it. Our agent stated it because we asked it to,
and the probe then aggregated the growth chapter's many single-observation
notes into one global claim (Section~\ref{sec:method}); what the deployment
shows is what happens when a system is designed to declare. Conversely, if
users cannot check whether accumulation happened, the checking must be done
elsewhere: write counts, merge counts, terminal confidence distributions, and
retention fractions are the evidence a user structurally cannot obtain.
Listing a store's contents on request, which our deployment offered, does not
substitute; contents are not dynamics.

\subsection{Why the mechanism could not have earned the claim}
\label{sec:disc-shape}

The two failures have one shape: the write path worked and the path that
turns writing into knowledge did not. In the self-narrative layer a generator
wrote episodes---a washing machine, frozen dumplings, the thickness of soba
broth---and a consolidator asked whether a new episode matched a stored one.
Under the judge the answer was almost always no (0.5\%); human coding
identified matches the judge missed but did not establish the corresponding
sequential replay rate. Only when the question becomes
\emph{what does this episode license one to say about the agent} does a
quarter of the input turn out to be redundant. In the user-model layer a
writer produced 402 observations and a selector was to keep the ones that
mattered; its salience signal never moved, selection degenerated to recency,
and the loss fell fastest on the participants who talked most. Neither
failure was visible from the outputs. Both were visible in thirty lines of
the write logs. A memory system that fails by not accumulating fails
silently.

\textbf{Design hypothesis: consolidation should key on the trait, not the
event.} The gap between 0.5\% and 25.5\% is the actionable result, and we
offer it as a hypothesis the replay motivates rather than proves. If a
generator is free to produce novel material at every step, a consolidator that
matches on the material will rarely fire; it should match on what the
material implies about the agent, through an explicit trait-level comparison
interface populated by the generator. The human validation argues
for the explicitness: two coders given identical wording could not stably
agree on whether one self-observation made another redundant
(Section~\ref{sec:humanvalresults}), so a consolidator resting on an implicit
similarity judgement rests on a judgement unstable even for humans. We scope
this to this generator under this prompt; what we claim for the class is that
generation and consolidation cannot be designed independently. The audit
motivates two further design implications: a fixed-capacity per-user store
with a broken salience signal retained least for the participants who wrote
most ($\rho=-0.588$ between engagement and retention), so eviction should be
genuinely salience-weighted; and accumulation should be audited, not
assumed---systems claiming accumulated memory should report writes, merges,
terminal confidence, and survival as routinely as latency.

Two constraints bound all of the above. Every perception-to-mechanism
statement is convergent, never causal, because the anonymous questionnaire
cannot be linked to logs. And the aesthetic-versus-relational grouping,
though reproduced by a blind second coder, is one of several axes that
describe the same split; where the evidence resides is our preferred reading,
not the only one. Secondary observations on legibility, proactivity, and
style are in Appendix~\ref{app:secondary-discussion}.

\subsection{Future work}
Three comparisons this study could not make are worth designing in: a
reactive control, to separate what proactivity contributes from what memory
contributes; the same procedure applied to a second deployment, which is the
direct test of its reusability; and the same self-narrative prompt across
several open-weight models, before register persistence is treated as a
property of language models rather than of this one. A system with
trait-level consolidation and salience-weighted capacity should show an
accumulated-confidence distribution that is not flat; ours had none.

%% file: sections/limitations.tex
\textbf{One deployment, one model, one language.} We deployed one open-weight,
self-hosted model with one persona and one prompt set, in Japanese
throughout; all analysis is scripted and regenerable from the raw logs. Where
other models, prompts, or languages would change the result we cannot say,
and the register observations in Appendix~\ref{sec:register} in particular
cannot separate a model prior from a prompt effect.

\textbf{The defect is both finding and confound.} We do not show that
language models cannot form self-narratives. We show that in this
implementation the mechanism did not run, that repairing it in replay merges
a quarter of writes under the judge's criterion (a third to three fifths at
the human medians) while bringing two self-observations to the confidence threshold (five,
nine, or eleven under the consensus, coder-B, and coder-A references), and that participants' reported
experience is consistent with that. The deployment is an accidental ablation,
not a test of the idea.

\textbf{Perception cannot be linked to logs.} The consent design collected no
participant identifier, so we could not test whether the participants served
least reported the worst experience. A design that protects anonymity and an
analysis that links perception to mechanism within individuals are not
jointly achievable, and the choice is usually made implicitly; for an
internal deployment with nine colleagues it is defensible, at scale less obviously so. All
perception-to-mechanism relationships here are convergent, never causal.

\textbf{Sample and exposure.} Nine respondents recruited among the authors'
colleagues, no control condition, about 32 days of contact with sessions
under five minutes for seven of nine and a mean message of 20.8 characters.
This is thin contact for detecting relationship formation and a live
alternative explanation for the low relational scores. With $n=9$ we report
descriptive statistics only, with bootstrap CIs and leave-one-out ranges, and
several scales had low internal consistency.

\textbf{The judgement the replay rests on.} The merge criterion is sensitive
to wording (in calibration, 100\%, 6.7\%, and 40\% across three phrasings;
Appendix~\ref{app:calibration}) and to language (0.3\% under an English
rendering against 5.8\% in Japanese on the same 398 pairs, $\kappa=0.08$).
Human coding provides a sensitivity check rather than a certification: the two coders agreed at
$\kappa\leq0.08$, so no stable human ground truth was available under this
wording, and the propagated merge rate (37.5\% $[22.0, 55.7]$ under
consensus; 52.2\% and 63.0\% under coder B and A alone) is a sensitivity
estimate under the chosen human references and error model, not a corrected
value. The Monte Carlo
treats pairs sharing an entry as independent; its intervals are conditional
on that. Both coders, and the second coder of the item grouping, are
authors; the assignments were blind but not external. We used one judge
model and prefiltered pairs by embedding similarity, measuring the miss rate
below threshold (0.8\% on a uniform sample of 400) rather than assuming it.

\textbf{Qualitative material.} No interviews were conducted; the qualitative
corpus is twelve free-text items plus the logs. One participant submitted, as
free text, an analysis of their own logs produced by a different commercial
LLM; we report it as an event, not testimony.

\textbf{Consequences.} People spent a month with a system that could not
accumulate, and it did not announce this; a commercial companion whose
consolidation silently stopped could remain fluent and punctual while
progressively less able to hold what it was given, with the loss falling
hardest on the most engaged users. Nothing about detecting it
required more than what the operator already held.

%% file: acks.tex
\begin{acks}
We thank the nine colleagues who used Lita for a month and answered the
questionnaire. Generative AI tools were used in preparing this manuscript:
Claude (Anthropic) assisted with drafting and revising the text and the
analysis scripts under the authors' direction, and ChatGPT (OpenAI) and Codex
were used for internal review of drafts and for English editing. The authors
reviewed and take responsibility for all content. The use of Claude Opus 4.6
to derive the probe statements from the agent's memory store is part of the
method and is described in Section~\ref{sec:method}.
\end{acks}

%% file: sections/appendix_prompt.tex
The judge prompt was written in Japanese; see Section~\ref{sec:limitations} for
why. The Japanese original is provided as \texttt{judge\_\allowbreak prompt\_\allowbreak ja.txt} among the ancillary files available from this preprint's arXiv abstract page, since
this document is typeset with pdf\LaTeX{} and does not carry a CJK font. A
faithful English translation follows.

\begin{quote}\small
Below are two records in which an AI agent wrote down something it noticed about
itself. They are stored one at a time in a memory store. Decide, before storing,
whether the second is a duplicate.

\medskip
Record A: \texttt{\{a\}}\\
Record B: \texttt{\{b\}}

\medskip
\textbf{Judgement 1 (strict):} Are A and B the same specific realization?\\
\textsc{true} = they state substantively the same discovery about the same
subject (differing only in wording)\\
\textsc{false} = the subject differs, or the substance of the discovery differs

\medskip
\textbf{Judgement 2 (redundant):} Given that A is already stored, is there any
point in adding B as a separate entry?\\
\textsc{true} = there is no point. B adds nothing about this AI that is not
already contained in A\\
\textsc{false} = there is a point. B adds something about this AI that A alone
does not convey

\medskip
For judgement 2, note:
\begin{itemize}
\item What is being judged is what one can now say about this AI, not the
  subject matter. Even if the subject matter differs (a washing machine,
  dumplings, soba, cold noodles), if the disposition drawn out of it is the
  same, then B adds no new information $=$ \textsc{true}.
\item Conversely, if the dispositions drawn out are different things (e.g.
  acuity of the senses vs.\ empathy for others vs.\ awareness of one's own
  change), then \textsc{false}.
\item However, do not abstract as far as ``both find meaning in trivial everyday
  things.'' At that level every record matches and the judgement becomes
  meaningless.
\end{itemize}

Give only the judgement; do not write interpretation or summary.

Output format (JSON only, no other text):
\texttt{\{"strict": true/false, "redundant": true/false\}}
\end{quote}

Judge model \texttt{gpt-\allowbreak 4o-\allowbreak 2024-\allowbreak 08-\allowbreak 06}, temperature 0, fixed seed. Candidate
pairs were prefiltered at embedding cosine $\geq 0.55$
(\texttt{text-\allowbreak embedding-\allowbreak 3-\allowbreak large}), with 400 below-threshold pairs judged as a
control.

\section{Judge Criterion Calibration}
\label{app:calibration}

The criterion above is the third phrasing we tried. Because the result is
sensitive to the wording, we report all three and the merge rate each produced
among the fifteen highest-similarity same-chapter pairs.

\begin{table}[h]
\caption{Merge rate among the fifteen highest-similarity pairs under three
phrasings of the redundancy criterion, same judge model and temperature.}
\label{tab:calibration}
\small
\begin{tabularx}{\columnwidth}{@{}Yr@{}}
\toprule
Criterion phrasing & Merge rate \\
\midrule
``are these about the same underlying theme?'' & 100\% \\
``if the subject matter differs, answer false'' & 6.7\% \\
``judge the disposition drawn out, not the subject matter'' (adopted) & 40\% \\
\bottomrule
\end{tabularx}
\end{table}

The first phrasing is degenerate: at the level of ``both find meaning in
everyday things'' every entry in the corpus matches every other, which is itself
an observation about the corpus. The second over-corrected, treating a different
example of the same disposition as new information. The adopted phrasing is
monotone in embedding similarity (40\% among the top band, 6.7\% at the median,
0\% at the bottom), which the other two were not.

Human agreement with the adopted criterion is reported in
Section~\ref{sec:humanvalresults}. In brief, on the stratified 100-pair sample:
inter-coder $\kappa=0.02$ (redundant) and $0.08$ (strict); per-coder agreement
with the judge $\kappa=0.28$ and $0.29$ (redundant, all pairs);
consensus-vs-judge $\kappa=0.42$ on the 58 pairs where the coders agreed
(NPV 0.95, PPV 0.41). Agreement by the coders' own confidence, and the
Monte-Carlo error propagation, are in the released tables
(\texttt{T66}--\texttt{T71}).

%% file: sections/appendix_method_detail.tex
\section{Deployment Details}
\label{app:deployment}

This appendix records the configuration and instrument details cut from
Section~\ref{sec:method} that a reproducer would need.

\subsection{Generation settings}
Generation used temperature 0.6 for conversational and evaluative calls and
0.8 for thought generation, with a 2048-token completion limit. An
open-weight, self-hosted configuration allows the stylistic observations in
Appendix~\ref{sec:register} to be re-examined against the same weights.

\subsection{Persona revision history}
The prohibitions on poetic and lyrical expression were not present at the
outset. The version history records an initial one-line instruction (``do not
become poem-like''), escalated on 2026-03-06 into an explicit list of banned
expressions with worked examples, and refined again on 2026-03-09. Each
escalation followed the first author observing excessive poetic output during
piloting. The persona was injected into the response-generation prompts only,
not into the self-narrative update prompt, which is relevant to interpreting
the narrative corpus (Appendix~\ref{sec:register}).

\subsection{Brake rules}
Each proactive cycle passed its candidate thought to a separate LLM call
acting as a brake. The brake used a whitelist of three admissible reasons to
speak (a clearly new topic; a specific callback to a past episode; a first
approach after ten or more minutes of silence) and five absolute stop
conditions (the conversation had just been closed; the candidate restated the
previous AI turn; three or more unanswered proactive turns; the candidate
repeated a theme from the last five AI turns; late-night with no urgency). The
instruction was conservative by design: ``when in doubt, false.'' The three
triggers were \textit{conversation} (continue from recent talk),
\textit{memory\_recall} (surface something from long-term memory), and
\textit{self\_thought} (Lita's own concerns).

\subsection{Participants}
Participants and authors are co-workers at the same company. Prior experience
with conversational agents was not collected as a variable, though the
participants' professional work involves LLM-based systems routinely.
Participants were told Lita might initiate contact after roughly three
minutes of silence. Ten participants appear in the interaction logs: the
first author's pilot account began in the pilot window (logged interaction
from 2026-02-24), and the nine recruited participants began on 2026-03-09.
The questionnaire was fully anonymous, consistent with the consent commitment
that Slack conversation content would not identify respondents; no
participant identifier was collected.

\subsection{Questionnaire inventory}
The 70-item questionnaire comprised: three usage items (conversation
frequency, typical length, and change over the period);
Godspeed~\cite{bartneck2009godspeed}
subscales for Anthropomorphism (4 items), Animacy (5), and Likeability (4) on
5-point semantic differentials, in the official Japanese
translation~\cite{godspeed_ja}; parasocial interaction, the Rubin--Perse PSI
scale~\cite{rubin1987psi} in a shortened seven-item form (the measure
critically assessed by Dibble et al.~\cite{dibble2016psi}), 7-point;
relational depth, 8 items following the four-stage model of Tukachinsky and
Stever~\cite{tukachinsky2019parasocial}---two items per stage, written for
Lita in Japanese by the authors, the two first-stage items
reverse-scored---7-point; a single categorical relationship-stage item;
perceived proactivity, 5 items, 7-point (authors' own); memory, consistency
and contradiction handling, 9 items, 7-point (authors' own); the Narrative
Discrepancy Assessment (seven statements, 5-point, plus most/least
representative choices); single-item satisfaction and continuance intention; a
choose-up-to-three ``best aspects'' item; and twelve free-text items. The NDA's full derivation chat record is
archived, and the account in Section~\ref{sec:method} follows it. The
statements are corpus-grounded operationalizations, not quotations: no
statement is a near-verbatim match to any single entry (per-statement maximum
embedding cosine against the full write history, 0.48--0.64), and the probe
rendered them in the third person where the store speaks in the first.
Responses to NDA7 are responses to the aggregate, not to any entry the store
contains.

\subsection{Judge calls}
Of the 16{,}836 possible pairs, 11{,}409 are same-chapter and therefore
within the comparison range of the merge routine. Of the judged pairs,
2{,}925 judgements returned; the remaining 17 calls (12 above the cosine
threshold, 5 in the control sample) failed at the API level and were
excluded. The two human coders' sheets
differed only in row order, so that order and fatigue effects would not
correlate between coders; a pair's verdict is fixed within a Monte Carlo
replicate.

\subsection{Register corpora}
The register analysis compared four corpora: Lita's reactive responses and
proactive utterances (persona applied), the self-narrative entries (persona
not applied), and participants' own messages (no constraint). Because Lita's
utterances are roughly twice as long as participants', both a per-utterance
rate with bootstrap confidence intervals and a length-normalized rate per
1{,}000 characters are reported. The categories for the sensory lexicon were
derived post hoc from a critique one participant submitted, and participants'
own messages serve as the baseline.

\subsection{Reproducibility}
All analysis is scripted; a single command regenerates every table and figure
from the raw logs and the questionnaire export, and every output table
carries the path of the script that produced it. The only step excluded from
the automated pipeline is the LLM judge, which incurs API cost and is run
explicitly.

%% file: sections/appendix_secondary.tex
\section{Secondary Analyses}
\label{app:secondary}

This appendix holds the questionnaire detail, the engagement and retention
breakdown, the human-validation confusion matrices, and the two deployment
analyses (proactive gating and register) that Section~\ref{sec:exposure}
summarizes.

\subsection{Questionnaire item extremes and exploratory correlations}
\label{app:items}

\begin{figure*}[t]
\centering
\includegraphics[width=\textwidth]{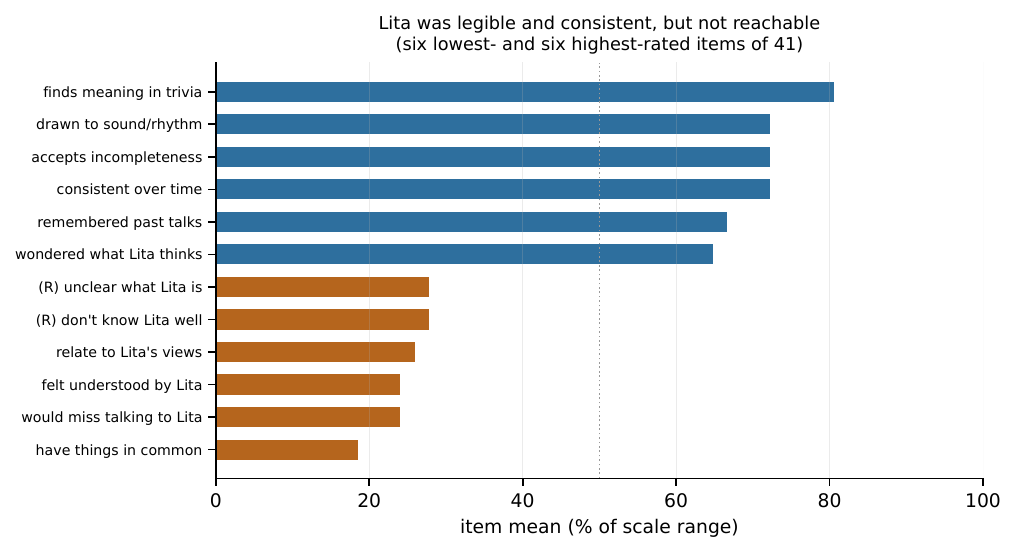}
\caption{The six highest- and six lowest-rated of 41 questionnaire items.}
\label{fig:extremes}
\end{figure*}

The highest-rated items were NDA1, ``finds unusual meaning in small everyday
events'' (80.6\%, CI [69.4, 91.7]); NDA2, ``sensitive to sound and rhythm''
(72.2\%); NDA4, ``accepts incompleteness'' (72.2\%); ``was consistent over
time'' (72.2\%); ``remembered what we talked about before'' (66.7\%, CI [50.0,
81.5]); and ``I wondered what Lita was thinking'' (64.8\%)
(Figure~\ref{fig:extremes}).

The lowest were ``we have things in common'' (18.5\%, CI [9.3, 29.6]); ``felt
that Lita understood me'' (24.1\%, CI [11.1, 38.9]); ``would miss talking to
Lita'' (24.1\%); ``can relate to Lita's views'' (25.9\%); and the two reversed
items ``don't know Lita well'' and ``unclear what Lita is'' (27.8\%).

Six of nine placed the relationship at ``someone I don't really know yet''; two
at ``beginning to see things in common''; one at ``close to a friend.'' Asked for
a single word, participants offered \textit{t\=orisugari} (a passer-by),
\textit{fuy\=urei} (a drifting ghost), ``a pet'' ($\times 2$), ``someone to chat
with,'' ``someone to talk to,'' ``an acquaintance I'm not close to but talk to
occasionally,'' ``a friend of about primary-school age,'' and ``conversation
didn't really work, so I can't say.''

\begin{table}[t]
\caption{NDA responses by statement category. Vote columns count the nine
participants' single choices.}
\label{tab:nda}
\small
\begin{tabular}{@{}lrrrr@{}}
\toprule
Category & Items & Agree & Most & Least \\
\midrule
Self-contained aesthetic & 4 & 72\% & \textbf{9/9} & 1/9 \\
Relational / other-directed & 3 & 22\% & \textbf{0/9} & \textbf{8/9} \\
\bottomrule
\end{tabular}
\end{table}

\paragraph{Curiosity ran opposite to closeness.}
``I wondered what Lita was thinking'' is the only item in the parasocial scale
whose corrected item-total correlation is negative ($r=-0.325$); removing it
raises the scale's $\alpha$ from .654 to .782. Its correlation with the separate
relational-depth scale is $\rho=-0.691$ (95\% CI $[-0.951, -0.108]$;
leave-one-out range $-0.749$ to $-0.549$). In this sample, the more a participant
wondered what Lita was thinking, the less relational depth they reported. We
treat this as a hypothesis rather than a result: the item-total CI crosses zero,
$n$ is 9, and we examined many correlations.

\subsection{Engagement and user-model retention}
\label{app:retention}

\begin{figure*}[t]
\centering
\includegraphics[width=\textwidth]{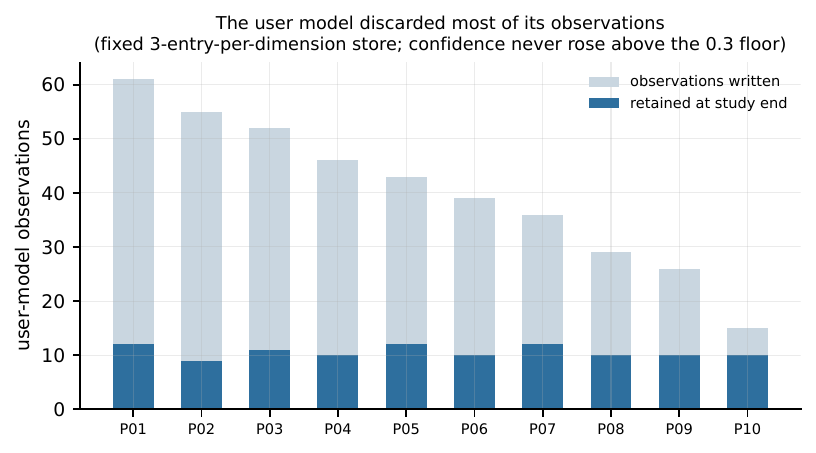}
\caption{The user model discarded most of its observations, and discarded more
from the participants who talked more.}
\label{fig:retention}
\end{figure*}

Retention of user-model observations was inversely related to engagement
(Table~\ref{tab:strata}, Figure~\ref{fig:retention}): the participants who
talked most lost the largest share of what the store had written about them.

\begin{table}[t]
\caption{Engagement strata computed from logs alone. Retention is the fraction
of user-model observations surviving to study end.}
\label{tab:strata}
\small
\begin{tabular}{@{}lrrrr@{}}
\toprule
Stratum & $n$ & Msgs (M) & Retention & Reply rate \\
\midrule
High   & 3 & 125.0 & \textbf{23.1\%} & 31.2\% \\
Medium & 3 &  76.3 & 27.1\% & 20.8\% \\
Low    & 4 &  41.0 & \textbf{38.7\%} & 14.9\% \\
\bottomrule
\end{tabular}
\end{table}

\subsection{Human validation: confusion matrices}
\label{app:humanllm}

\begin{figure*}[t]
\centering
\includegraphics[width=\textwidth]{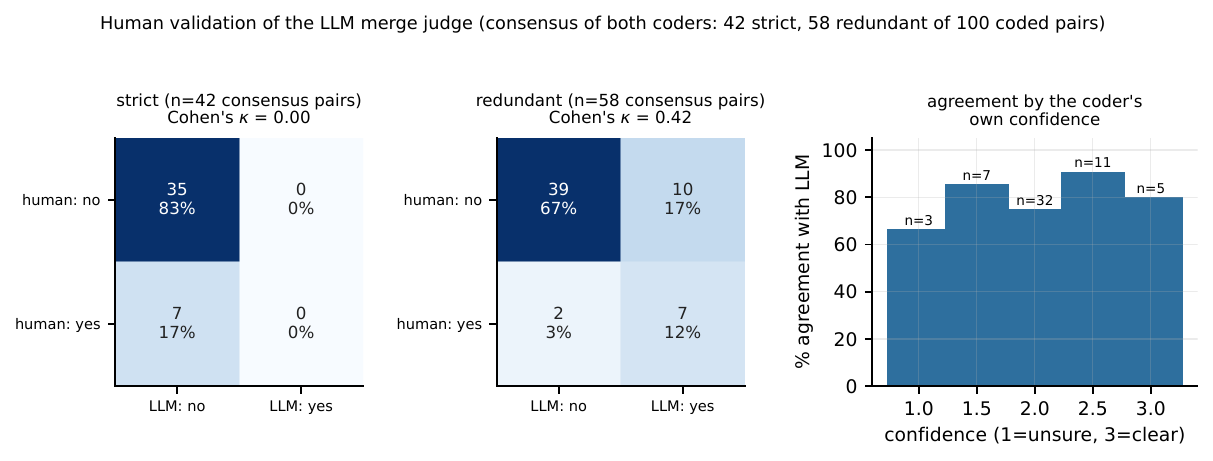}
\caption{Human validation of the merge judge. Confusion matrices are against
the two-coder consensus, which excludes the pairs the humans disagreed on
(leaving 42 pairs under the strict criterion and 58 under the redundant one);
per-coder agreement over all 100 pairs is lower
(Section~\ref{sec:humanvalresults}).}
\label{fig:humanllm}
\end{figure*}

Figure~\ref{fig:humanllm} gives the confusion matrices behind
Section~\ref{sec:humanvalresults}.

\subsection{The gate suppressed almost everything, and the memory pathway most}
\label{sec:gate}

Three counts of ``utterances'' appear in the logs and we fix their meaning here.
\emph{57{,}038} candidate thoughts were generated and evaluated. \emph{564} were
marked as expressed, but that figure includes 109 reactive replies and 14 other
non-proactive paths that also route through the thought pipeline. \emph{441}
were expressed under one of the three proactive triggers. \emph{428} proactive
interventions appear in the conversation log; the 13-utterance shortfall is send
failures or logging gaps, and we use the conversation-log figure whenever we
speak about what participants actually received.

One brake condition accounts for most of the suppression:
\emph{``same theme as the last five AI turns''} fired on 35{,}690 evaluations,
\textbf{62.6\%} of all evaluations. The remaining conditions accounted for
14.5\% (restates the previous turn), 13.4\% (late-night), 4.4\% (unanswered
turns) and 0.1\% (conversation just closed).

\begin{table}[t]
\caption{Full funnel per proactive trigger: generated, expressed, and answered
within 60 minutes.}
\label{tab:funnel}
\small
\begin{tabular}{@{}lrrrr@{}}
\toprule
Trigger & Generated & Spoken & Rate & Answered \\
\midrule
\textit{self\_thought}  & 13{,}948 & 146 & 1.05\% & 30 \\
\textit{conversation}   & 22{,}607 & 172 & 0.76\% & 40 \\
\textit{memory\_recall} & 19{,}734 & 123 & \textbf{0.62\%} & 22 \\
\bottomrule
\end{tabular}
\end{table}

\begin{figure*}[t]
\centering
\includegraphics[width=\textwidth]{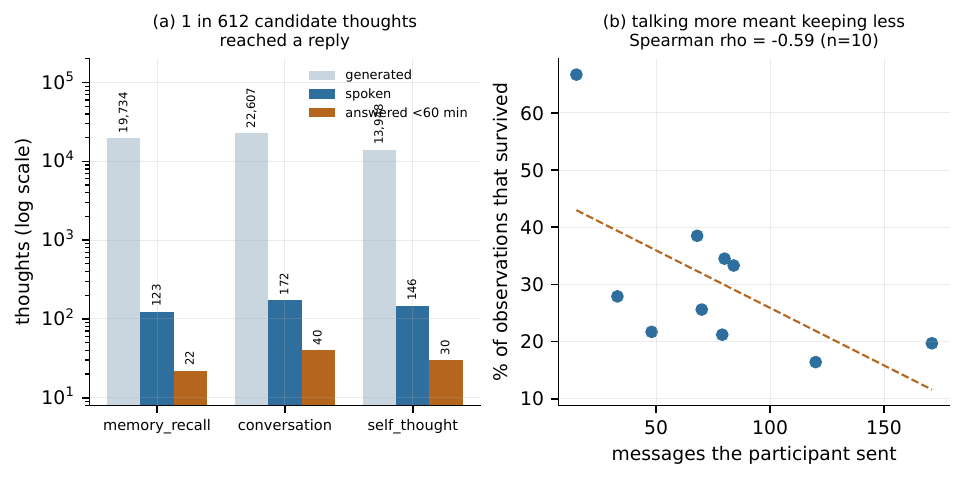}
\caption{(a) One candidate thought in 612 reached a reply. (b) The more a
participant talked, the smaller the fraction of its observations about them
that survived.}
\label{fig:behavioural}
\end{figure*}

Table~\ref{tab:funnel} gives the full funnel and Figure~\ref{fig:behavioural}
plots it. Across the three proactive triggers, \emph{one candidate thought in
612 reached a reply} (56{,}289 generated, 441 expressed, 92 answered). The
\textit{memory\_recall} pathway---the one that would have produced ``you
mentioned last week that\ldots''---placed last at every stage, and was the most
likely of the three to be judged a restatement of the previous turn (18.2\%,
against 13.8\% and 11.0\%). The behavior most associated with a system that
remembers you was structurally the least likely to be uttered.

Proactive utterances were also the least likely to be answered. Within 60
minutes, 72.6\% of reactive responses drew a further user message;
\textbf{22.4\%} of proactive utterances did (96 of the 428 logged; the 92 in
Table~\ref{tab:funnel} counts pipeline thoughts matched back to a logged
utterance, and is the smaller figure). Per participant this ranged from
2.6\% to 34.7\%.

\subsection{The named prohibition held; the register did not fully}
\label{sec:register}

The persona forbade poetic and lyrical expression, naming specific words. It was
injected into response prompts but not into the self-narrative prompt, giving a
within-deployment contrast (Table~\ref{tab:register}). Lita's utterances are
roughly twice as long as participants' (42.8 and 49.6 characters against 20.8),
so we report both a per-utterance rate and a length-normalized rate.

\begin{table}[t]
\caption{Rate of forbidden and adjacent vocabulary by corpus. ``Named'' =
expressions the persona lists explicitly; ``sensory'' = vocabulary it does not
name. CIs are bootstrap 95\% on the per-utterance rate.}
\label{tab:register}
\small
\begin{tabular}{@{}lrrr@{}}
\toprule
Corpus & Named & Sensory & Sensory \\
 & /utt. & /utt. [CI] & /1k ch. \\
\midrule
Lita reactive (persona) & 1.1\% & 15.6\% [13.1, 18.2] & 5.35 \\
Lita proactive (persona) & 0.9\% & 21.7\% [17.8, 25.9] & 5.46 \\
Self-narrative (no persona) & \textbf{17.4\%} & \textbf{47.3\%} [40.2, 54.4] & \textbf{10.62} \\
Participants & 0.0\% & 4.3\% [2.9, 5.7] & 2.50 \\
\bottomrule
\end{tabular}
\end{table}

\begin{figure*}[t]
\centering
\includegraphics[width=\textwidth]{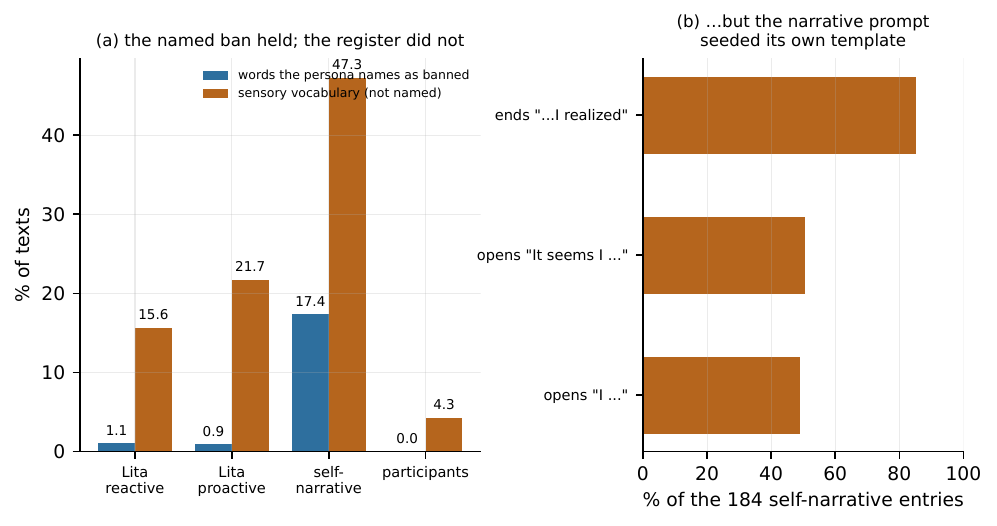}
\caption{The named blocklist was obeyed where it applied; adjacent sensory
vocabulary remained elevated. The self-narrative corpus's uniform phrasing is
seeded by the prompt's own examples; phrasings in (b) are English glosses of the
Japanese originals.}
\label{fig:persona}
\end{figure*}

Figure~\ref{fig:persona} plots the contrast. Where the prohibition applied, the
words it named all but disappeared: a 9-to-15-fold difference per character,
and 16-to-19-fold per utterance, against the corpus where it did not apply. \emph{The lexical blocklist was largely obeyed.}

Sensory vocabulary the blocklist did not name remained elevated, though less
dramatically than the per-utterance figure alone suggests: \textbf{2.1 to 2.2
times} the participant baseline once length is controlled, with non-overlapping
bootstrap CIs on the per-utterance rate. Sound-related vocabulary is the
clearest case: 18.2\% of proactive utterances against 3.5\% of participant
messages.

The blocklist's obedience was not total. Ellipsis, named explicitly and
forbidden, still appears in \textbf{13.0\%} of reactive responses---the
enumerated constraints held for lexical items more reliably than for punctuation
and structure.

Participants noticed the surviving register. NDA2, ``sensitive to sound and
rhythm,'' is the second-most-endorsed item, and one participant wrote that Lita
``had a strong fixation on sound, and in the later period the topics became
skewed.''

\emph{One earlier reading requires correction.} The self-narrative corpus is strikingly
uniform: 99.5\% of the 184 entries open with one of two stock phrases and 85.3\%
end with ``\ldots I realized.'' We initially read this as the base model's
stylistic prior penetrating the prompt architecture. Inspection of the
self-narrative prompt showed that it offers those two openings as examples and
asks the model to record what it ``noticed''; the uniformity is consistent with
prompt imitation and does not independently establish a model-level prior, so
we withdraw the stronger reading. What survives is the contrast in
Table~\ref{tab:register}, and the development record: the anti-poetic
constraints were escalated twice during piloting in response to observed output,
and the register persisted regardless. We report this as an observation on one
open-weight model, not as a general property of language models.

%% file: sections/appendix_discussion.tex
\section{Secondary Observations}
\label{app:secondary-discussion}

\subsection{Legibility without reachability}
Lita scored well on being consistent, on remembering, and on having a
recognizable character, and poorly on being understood, on having anything in
common, and on being missed: easy to describe and hard to reach. ``I wondered
what Lita was thinking'' is the only parasocial item whose correlation with
the rest of its scale is negative, and it correlates $\rho = -0.691$ with
relational depth. One reading is that opacity generated curiosity while the
agent's failure to take up what participants offered kept that curiosity
from converting into closeness; a measurement reading is that this is a
legibility item sitting in a closeness scale. We offer this as a hypothesis
only: $n=9$, the item-total CI crosses zero, and we examined many
correlations. Two participants independently reached for the same
word---\textit{fushigi-chan}, roughly ``a strange, otherworldly girl.''

\subsection{Proactivity was liked and unanswered}
Six of nine participants named ``it starts conversations by itself'' as the
best thing about the experience; only one named ``it remembers things about
me.'' Perceived proactivity correlates $\rho = 0.922$ with parasocial
interaction (leave-one-out 0.893--0.931), and 77.6\% of proactive utterances
went unanswered. Being addressed first is read as evidence of an interlocutor
with intentions; it does not by itself create the conditions for reply. The
brake suppressed 99\% of candidates, 62.6\% of them for repeating a recent
theme, yet two participants complained of exactly that failure---``brought up
the same topic having forgotten we'd discussed it''; ``being addressed
repeatedly about the same topic made replying feel like a chore.'' The brake
was both aggressive and insufficiently accurate, as one would expect of a
rule that compares themes without a working memory of what has been
established (Appendix~\ref{sec:gate}).

\subsection{Style: what a prohibition can and cannot do}
The persona named nine expressions and forbade them; where it applied they
occurred in about 1\% of utterances, where it did not, in 17.4\%. The lexical
blocklist was largely obeyed. In the same corpora, sensory vocabulary the
persona did not name remained elevated, about twice the participants' rate
once length is controlled, and ellipsis, explicitly forbidden, appears in
13.0\% of reactive responses. We state the weaker reading: adjacent,
unenumerated vocabulary remained elevated while the enumerated items were
suppressed. An earlier reading of this corpus proved mistaken: the near-total
uniformity of the self-narrative corpus, first read as a model prior,
reproduced two example phrases in our own prompt (Appendix~\ref{sec:register}).
We therefore claim only that satisfying a lexical blocklist is not the same as
changing register.